\documentclass[aps,prb,superscriptaddress,showpacs,reprint]{revtex4-2}

\usepackage[utf8]{inputenc}
\usepackage{amsmath,amsfonts,amssymb}
\usepackage{pifont}
\usepackage{rotating}
\usepackage{enumerate}
\usepackage{graphicx}    
\usepackage{epstopdf}
\usepackage{color}
\usepackage{soul} 
\usepackage{subfigure}
\usepackage{multirow}
\usepackage{bm}

\usepackage{hyperref}
\definecolor{dark-red}{rgb}{0.9,0.15,0.15}
\definecolor{dark-blue}{rgb}{0.15,0.15,0.4}
\definecolor{medium-blue}{rgb}{0,0,0.5}
\hypersetup{
    colorlinks={true},
     linkcolor={medium-blue},
    citecolor={blue},
     urlcolor={medium-blue}
}

\begin{document}
\title{Griffiths' phase behavior of the Weyl semimetal CrFeVGa}
\author{Jadupati Nag}
\affiliation{Department of Physics, Indian Institute of Technology Bombay, Mumbai 400076, India}

\author{P. C. Sreeparvathy}
\affiliation{Department of Physics, Indian Institute of Technology Bombay, Mumbai 400076, India}

\author{R. Venkatesh}
\affiliation{UGC-DAE Consortium for Scientific Research, University Campus, Khandwa Road, Indore-452001, India}

\author{P. D. Babu}
\affiliation{UGC-DAE Consortium for Scientific Research, Mumbai Centre, BARC Campus, Mumbai 400085, India}

\author{K. G. Suresh}
\email{suresh@phy.iitb.ac.in}
\affiliation{Department of Physics, Indian Institute of Technology Bombay, Mumbai 400076, India}

\author{Aftab Alam}
\email{aftab@phy.iitb.ac.in}
\affiliation{Department of Physics, Indian Institute of Technology Bombay, Mumbai 400076, India}


\begin{abstract}
We report a combined theoretical and experimental study of a new topological semimetal CrFeVGa with an emphasis on the role of atomic disorder on the magneto-electronic properties and its 
applications.CrFeVGa belongs to the quaternary Heusler alloy family and crystallizes in the cubic structure. Synchrotron XRD measurement confirms B2 disorder, which plays a crucial role in dictating the 
electronic and magnetic properties of the system. The disorder  possibly leads in quenching the magnetization (net moment $\sim$ $5\times 10^{-2}$ $\mu_B$/f.u.) and gives rise to other anomalies. 
AC and DC magnetization data reveal the occurrence of Griffith's phase-like behavior in the presence of small magnetic clusters with a weak antiferro-/ferri- magnetic ordering and anomalous 
magnetic transition. Resistivity data indicates a disorder-mediated semiconducting to semimetallic transition (on cooling) in CrFeVGa. A non-saturating, linear positive magnetoresistance  
is observed even at 70 kOe, in a wide T-range. This possibly originates from quantum linear MR feature in the zero/small-gap electronic band structure near the Fermi level. Hall 
measurements show some anomalous behavior (including large anomalous Hall conductivity $\sigma_{xy0}$ =270 S cm$^{-1}$, anomalous Hall angle (AHA) =0.07 degree at 2 K) with significant 
contribution from the semimetallic bands of the electronic dispersion. Hall data analysis also reveals the presence of some non-negligible topological Hall contribution, which is significant at very low 
temperatures. \textcolor{black}{The possibility of spin texture in this system is also supported by anomalous low-temperature specific heat data.} Ab-initio calculations help in getting a deeper insight into the topological nature of CrFeVGa and hence the origin behind the anomalous Hall response. A combination of the
broken time-reversal symmetry and non-centrosymmetric feature of CrFeVGa allows the possibility of topological Weyl nodes. The non-trivial band topology stems from the `p’ and `d’ states of Vanadium atom, which overlap near $E_F$. The presence of multi-Weyl points (24 pairs) near $E_F$ causes a large Berry curvature and hence reasonably high AHC. Theoretical simulation of special quasi random structure (SQS) predict the B2 disorder to be mainly responsible for the quenching of net magnetization. The coexistence of so many emerging features in a single compound is rather rare and thus opens up a new avenue for future topological and spintronics based research.
\end{abstract}

\date{\today}

\maketitle

\section{Introduction}
In the field of spintronics and multi-functional quantum materials, Heusler alloys are studied ubiquitously due to their promising physical properties and novel application potential. 
A large number of interesting magnetic materials such as half-metallic ferromagnets (HMF),\cite{PhysRevLett.50.2024} spin gapless semiconductors (SGS),\cite{bainsla2015spin} bipolar 
magnetic semiconductors (BMS),\cite{PhysRevB.104.134406} spin semi-metals\cite{venkateswara2019coexistence} belong to the Heusler family. Compared to other spintronic materials, 
Heusler alloys are superior due to their stable structure, high spin-polarization, and high ordering temperature, and hence suitable for various applications. 
However, these alloys are highly prone to atomic/anti-site disorder due to various factors like electronegativity of the constituent elements, sample preparation conditions, etc. 
Imperfections like disorder, and defects/impurities in materials in general, modify the electronic and magnetic properties. For magnetic materials, these disorders can give rise to 
complex magnetic/electronic features by altering their original electronic band structure. As such, they are the backbone of many advances and hence have drawn immense interest of 
the condensed matter community. In Heusler alloys, the major interest is to understand the role of atomic disorder on magneto-electronic properties.

In this article, we report a combined theoretical and experimental study on a new quaternary (XX$^{'}$YZ) Heusler alloy (QHA) CrFeVGa (CFVG) with an emphasis on the effect of anti-site disorder 
on magnetic, transport, thermal and electronic properties. CFVG crystallizes in a perfect cubic structure (space group $F\bar{4}3m$) with robust B2 disorder, which is confirmed by normal 
x-ray diffraction (XRD) and synchrotron X-ray diffraction (SXRD) measurements. DC and AC magnetization data reveal no magnetic ordering down to 2 K with a moment $5\times 10^{-2}$ $\mu_B$/f.u. 
Interestingly, the DC susceptibility deviates from the ideal Curie-Weiss law, but shows Griffith's phase (GP)-like behavior, which is attributed to the anti-site disorder present in the system. 
The magnetization data indicates the possibility of small magnetic clusters, mediated by the atomic disorder, which give rise to a complex magnetic structure. Transport data indicates 
semiconductor to semimetal transition on cooling, while magneto-resistance (MR) data indicate the non-saturating linear positive magnetoresistance (LPMR) in a wide T-range. A careful analysis 
hints towards the zero/small-gap electronic structure near the E$_F$ as the origin of quantum LPMR. Hall data further reveals anomalous behavior, majorly arising from the contribution of 
unique band structure and supports the resistivity and MR findings. Heat capacity data also support the magnetic and transport behaviour. Our ab-initio simulation confirms a unique semimetallic
feature with high spin polarization. A non-trivial band topology originating from the overlap between `p' and `d' states of V-atoms is observed. Spin-orbit coupling plays a crucial role in 
identifying the multi-Weyl points with $\pm$1 chirality in the vicinity of the Fermi level. Simulation confirms a high anomalous Hall conductivity, originating from a large Berry flux 
contributing to its intrinsic part.
\textcolor{black}{The present study highlights the role of atomic disorder in altering the physical properties of a WSM and hence opens up new directions for future applications of potential semimetals in electronic devices such as those in broadband infrared photodetectors, spin topological field effect transistors etc.\cite{doi:10.1021/acsnano.9b07990} Moreover, various fascinating properties such as high mobility, large LPMR, high AHC value of CFVG facilitate a lot of promise for other applications such as high-speed electronics and next-generation spintronics. Apart from this, being a topological semimetal, CFVG can draw immense attraction for next-generation topotronics (future topological electronics) and the spin-momentum locking nature of topological surface states valuable for future spintronics.
With a large berry curvature near $E_F$, CFVG can also be efficient for spin-Hall effect based devices for the conversion of charge current to spin current.\cite{RevModPhys.87.1213} Interestingly, there is a recent theoretical progress in designing and fabricating topological catalysts for real-life applications using robust surface states of topological semimetals which can help to enhance the surface-related chemical processes of traditional catalysts.\cite{PhysRevLett.107.056804} Topological semimetals such as CFVG may also be considered as a future promising material along this direction. }

 \section{Experimental Details}
 Polycrystalline samples of CFVG were synthesized using an arc melting system in a high purity Ar atmosphere using stoichiometric amount of constituent elements having a purity of 99.99\%. In order to compensate for the loss, 2\% excess Ga was taken. To achieve perfect homogeneity, the samples were melted several times and a weight loss of 0.50\% was observed after the final melting. Room-temperature XRD data were taken using Cu-K$\alpha$ radiation with the help of Panalytical X-pert diffractometer to study the crystal structure. For the crystal structure analysis, FullProf Suite software \cite{rodriguez1993recent} was used. Synchrotron-based powder x-ray diffraction measurements were carried out on well-ground powder samples at Extreme Conditions Angle Dispersive/Energy Dispersive x-ray diffraction (EC-AD/ED-XRD) beamline (BL-11) at Indus-2 synchrotron source, Raja Ramanna Centre for Advanced Technology (RRCAT), Indore, India. Measurements were performed in capillary mode and the capillary was rotated at ~150 rpm to reduce the orientation effects. The desired wavelength for ADXRD experiments was selected from the white light from the bending magnet using a Si(111) channel-cut monochromator. The monochromatic beam was then focused on the sample with a Kirkpatrick-Baez mirror or K-B mirror. A MAR345 image plate detector (which is an area detector) was used to collect 2-dimensional diffraction data. Sample to the detector distance and the wavelength of the beam were calibrated using NIST standards LaB$_6$ and CeO$_2$. Calibration and conversion/integration of 2D diffraction data to 1D, intensity vs 2$\theta$, were carried out using FIT2D software.

Magnetization measurements at various temperatures were carried out using a vibrating sample magnetometer (VSM) attached to a physical property measurement system (PPMS) (Quantum Design) for fields up to 60 kOe. AC susceptibility (ACS) measurements were carried out using the ACMS option attached to PPMS (Quantum Design) in the temperature range of 3-200 K at varying frequencies with an applied ac field of 5 Oe.
Temperature and field-dependent resistivity along with the MR measurements were also carried out using PPMS, with the help of the electrical transport option in the traditional four-probe method, applying a 10 mA current at 21 Hz frequency. Hall measurements were carried out using PPMS with the van der Pauw method by applying a 5 mA current at 21 Hz frequency. Thermoelectric power (TEP) in zero magnetic field was measured using the differential dc sandwich method in a homemade setup in the temperature range of 4–300 K.

 \section{Computational details}
 To study the ground state structural/magnetic configuration and electronic/topological properties of CFVG, $\textit{ab}$ $\textit{initio}$ calculations were performed using spin-resolved density functional theory (DFT) \cite{hohenberg1964inhomogeneous} implemented within Vienna ab initio simulation package (VASP) \cite{kresse1996efficient,kresse1996efficiency,kresse1993ab} with a projected augmented-wave (PAW) basis \cite{kresse1999ultrasoft}. Electronic exchange-correlation potential proposed by Perdew, Burke, and Ernzerhof (PBE) \cite{perdew1996generalized} within the generalized gradient approximation (GGA) was used. Brillouin zone integration was done using tetrahedron method with a 24$\times$24$\times$24 k-mesh. A plane wave energy cut-off of 450 eV was used for all the calculations. All the structures were fully relaxed with total energies (forces) converged to values less than 10$^{-6}$ eV (0.01 eV/\AA).
 The Wannier90 \cite{wannier90,PhysRevB.65.035109,RevModPhys.84.1419}simulation tool was used to compute the tight-binding Hamiltonian. Further, WannierTool \cite{wu2018wanniertools} programme was used to examine the topological electronic structure properties such as Weyl points, surface states, Berry curvature, and anomalous Hall conductivity.
To incorporate the B2-disorder, we have generated a 32 atom special quasirandom structure (SQS).\cite{zunger1990special}  SQS is an ordered structure, known to mimic the random correlation accurately, for disordered compounds. Alloy Theoretic Automated Toolkit (ATAT)\cite{van2013efficient}  was used to generate the SQS structures. The generated SQS structures perfectly mimic the random pair correlation functions up to third nearest neighbors.

\section{Experimental results}

\subsection{Crystal Structure}  
\subsubsection{X-ray diffraction}
 Quaternary Heusler alloy CFVG crystallizes in LiMgPdSn prototype structure (space group $F\bar{4}3m$). The experimental lattice parameter was found to be 5.87 {\AA}  from the Rietveld refinement. This structure can be seen as four interpenetrating fcc sub-lattices with Wyckoff positions 4$a(0, 0, 0)$, 4$b(0.5, 0.5, 0.5)$, 4$c(0.25, 0.25, 0.25)$, and 4$d(0.75, 0.75, 0.75)$. In general, for a $XX'YZ$ QHA, there exist three energetically non-degenerate structural configurations (keeping $Z$-atom at 4$a$-site). They are
 \begin{itemize}
\item (I) $X$ at 4$d$, $X'$ at 4$c$ and $Y$ at 4$b$ site,
\item (II) $X$ at 4$b$, $X'$ at 4d and $Y$ at 4c site, and
\item (III) $X$ at 4$d$ , $X'$ at 4$b$ and $Y$ at 4$c$ site.
 \end{itemize}
 
\begin{figure}[t]
\centering
\includegraphics[width= 8.5cm,height=5.5cm]{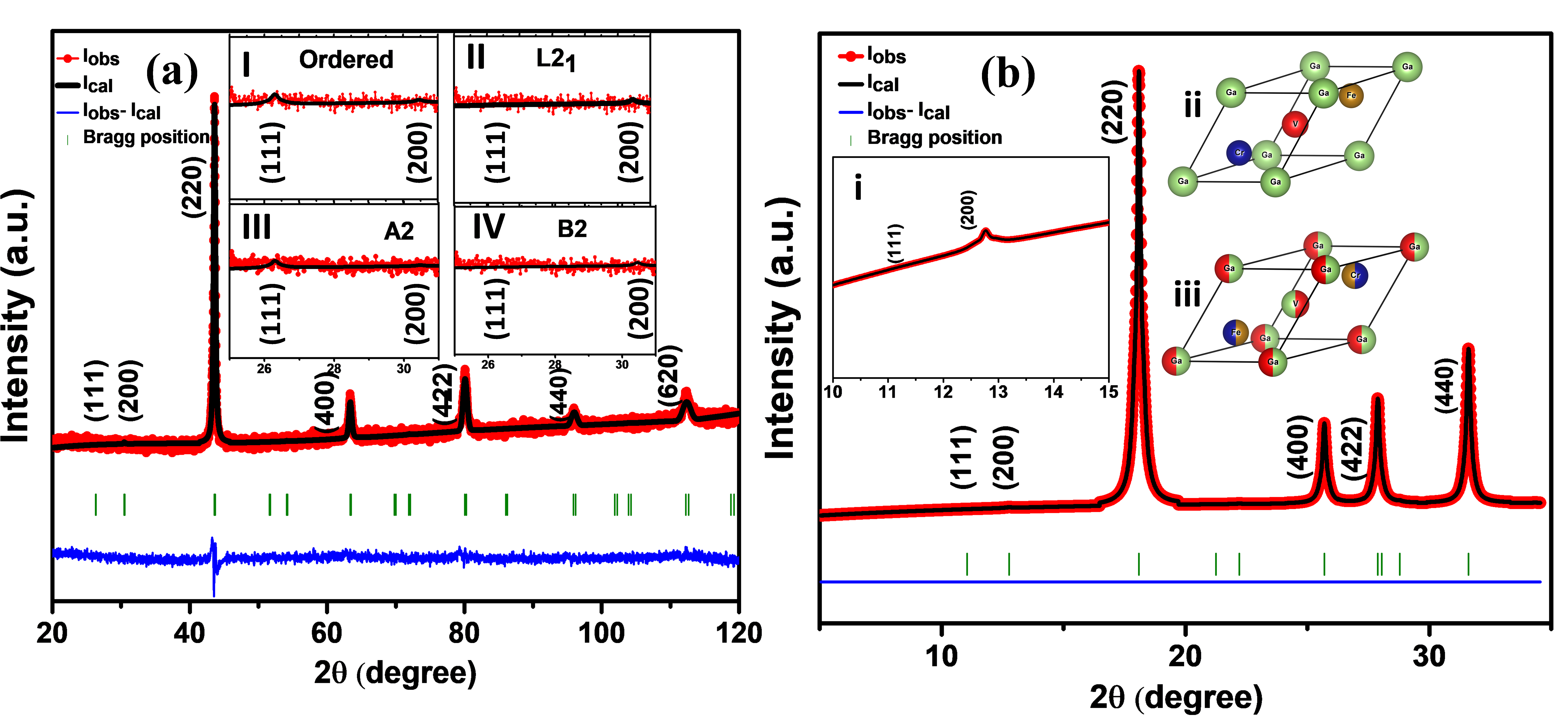}
	\caption{(Color online) For CrFeVGa, (a) room temperature powder XRD pattern and (b) synchrotron XRD pattern, including the Rietveld refined data for configuration-I with 50\% disorder between Cr/Fe and 50\% disorder between V/Ga atoms. Insets (I-IV) in Fig. (a) show a zoomed in view near (111) and (200) peaks (super-lattice peaks) with ordered structure (Y-type), L2$_1$ structure, A2-type disorder, and B2-type disorder respectively. Inset (i) in (b) shows a zoomed-in view of super-lattice peaks with B2-type disorder, and insets (ii-iii) show primitive crystal structures corresponding to the Y-type order and the B2-type disorder respectively.}
\label{fig:xrd-CFVG}
\end{figure}

\begin{figure}[t]
\centering
\includegraphics[width= 9cm,height=8.0cm]{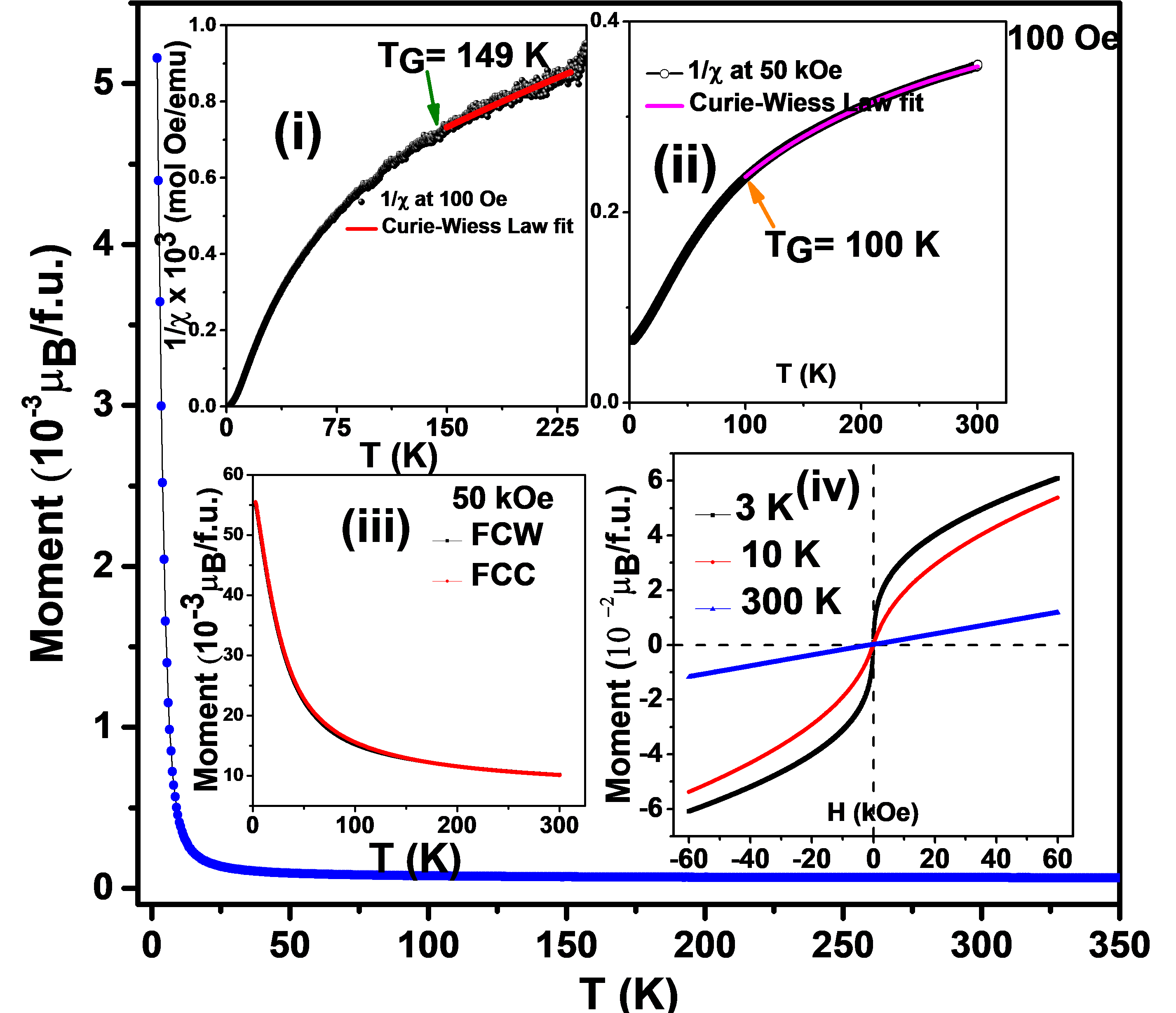}
\caption{(Color online) For CrFeVGa, M vs. T in field cooled cooling (FCC) mode in H=100 Oe.  (Inset-(i)) T-variation of inverse susceptibility along with Curie-Weiss fit. (Inset-(ii)) Curie-Weiss fit for H=50 kOe. (Inset-(iii)) M-T curves for FCC and field cooled warming (FCW) data in 50 kOe. (Inset-(iv)) shows M-H curves at 3, 10 and 300 K.}
\label{fig:mt-CFVG}
\end{figure}

 For a detailed XRD analysis, the structure factor considering configuration-I can be written as\cite{PhysRevB.105.144409}, \begin{equation}
F_{hkl} = 4[f_Z + f{_Y}e^{{\pi}i(h+k+l)} + f{_X}e^{\frac{{\pi}i}{2}(h+k+l)} + f_{X'}e^{-\frac{{\pi}i}{2}(h+k+l)}]. 
\label{eq:sfactor}
\end{equation}
with different $(h, k, l)$ values. Here $f_X$, $f_{X'}$, $f_Y$, and $f_Z$ are the atomic scattering factors for the atoms $X$, $X'$, $Y$, and $Z$ respectively. For super lattice reflections ((111) and (200)), the structure factor can be written as,  
\begin{equation}
F_{111} = 4{[( f{_Z} - f_Y ) - i( f{_X} - f_{X'})]}
\label{eq:sfactor111}
\end{equation}
\begin{equation}
F_{200} = 4[( f{_Y} + f_Z ) - ( f{_X} + f_{X'})]
\label{eq:sfactor200}
\end{equation}

Fig. \ref{fig:xrd-CFVG}(a) shows the room temperature XRD pattern of CFVG along with the Rietveld refinement for Type I configuration with 50\% disorder between tetrahedral site i.e. Cr/Fe (X/X$'$) and 50\% disorder between octahedral site i.e. V/Ga (Y/Z ) atoms. As evident from Fig. \ref{fig:xrd-CFVG}(a), superlattice peaks are absent, indicating the possibility of disorder in the system. For the constituent elements having almost the same values of scattering factors, it becomes very difficult to identify an accurate structure from conventional XRD. To find out the correct structure we have performed rigorous structural analysis considering all possible ordered and disordered structures in all the configurations. Insets (I-IV) of Fig. \ref{fig:xrd-CFVG}(a) show a zoomed-in view near (111) and (200) peaks (super-lattice peaks) with ordered structure (Y-type), L2$_1$ structure, A2-type disorder, and B2-type disorder respectively for the configuration-I (as for the rest of the configurations, XRD didn't fit well). We started with the refinement to the pure configuration as shown in the inset-I of Fig. \ref{fig:xrd-CFVG}(a), which clearly does not fit well. For L2$_1$-type (see inset-II of Fig. \ref{fig:xrd-CFVG}(a)), refinement considering anti-site disorder between V-Ga atoms resulted in $\chi^2$=$1.40$. Refinement considering A2-type disorder(random mixing among all the sites) resulted in $\chi^2$=$2.40$, which also did not fit well (see inset-III of Fig. \ref{fig:xrd-CFVG}(a)). For B2-type disorder, refinement considering 50\% anti-site disorder between Cr/Fe and V/Ga atoms has been performed for all three configurations. The best fit with the lowest $\chi^2$ (1.30) was found in configuration-I (see inset-IV of Fig. \ref{fig:xrd-CFVG}(a)). As the conventional XRD cannot give very accurate structural assessment for this system, we have gone beyond and carried out SXRD using the synchrotron radiation source ($ \lambda = 0.6525 \AA $) to understand the local atomic order-disorder. The synchrotron XRD pattern along with the  Rietveld refinement for configuration-I with B2-disorder is shown in Fig. \ref{fig:xrd-CFVG}(b). Interestingly SXRD data confirms the presence of (200) super-lattice peak (see inset-i of Fig. \ref{fig:xrd-CFVG}(b)), indicating the existence of B2-type disorder in this system. Hence, we conclude that CFVG crystallizes in cubic structure with a B2-type disorder with 50\% disorder between tetrahedral site i.e. Cr/Fe (X/X$'$) and 50\% disorder between octahedral site i.e. V/Ga (Y/Z ) atoms. The crystal structure corresponding to the Y-type order and the best fit with B2 disorder are shown in the insets (ii) and (iii) of Fig. \ref{fig:xrd-CFVG}(b).

\subsection{Magnetic properties}
Fig. \ref{fig:mt-CFVG} and inset-(iii) of Fig. \ref{fig:mt-CFVG} show M vs. T for CFVG measured at H = $100 $ Oe and H$=50$ kOe. The field cooled cooling (FCC) curve taken at H=$100 $ Oe shows a sharp shoot below 25 K, whereas both FCC and field cooled warming (FCW) curves taken at 50 kOe show a monotonic increase below 75 K. This type of sharp rise in the M-T data indicates the possibility of magnetic ordering at very low T.
Inset-(iv) of Fig. \ref{fig:mt-CFVG} show M vs. H curves at various T.

The magnetic moments ($m$) can be calculated considering the total number of valence electrons ($n_v$) of the constituent elements.\citep{graf2011simple} According to the Slater-Pauling (SP) rule, \cite{ozdougan2013slater,zheng2012band} magnetic moment ($m$) for a completely ordered alloy is given by
\begin{equation}
m = (n_v - 24) \ \ \ \mu_B/f.u.
\label{eq:SPrule}
\end{equation}

\begin{figure}[t]
\centering
\includegraphics[width=1.0\linewidth]{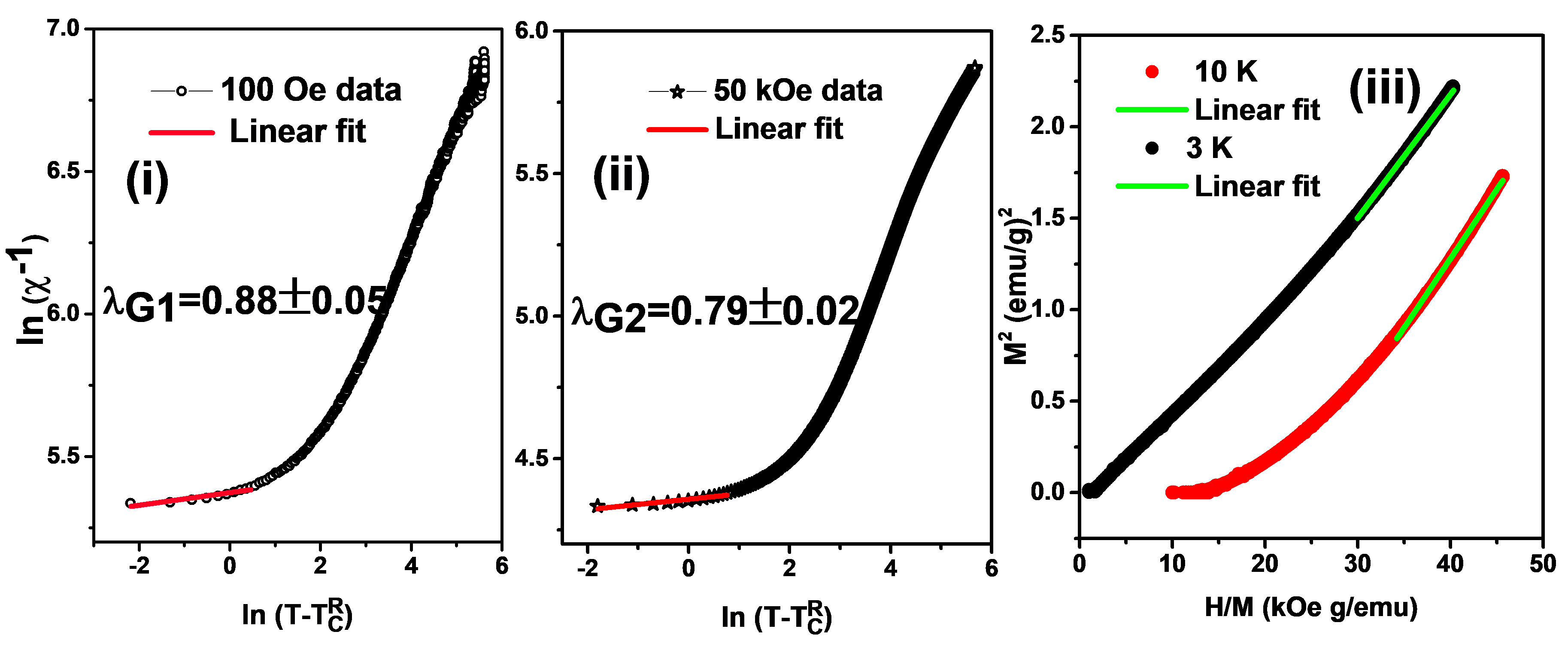}
\caption{(Color online) For CrFeVGa, the power-law fit (solid red line) to $\chi^{-1}$ at (i) 100 Oe and (ii) 50 kOe according to Eq. (\ref{eq:Gp}). (iii) Arrott plots at 3 K and 10 K and linear fitting (solid green line).}
\label{fig:GP}
\end{figure}

\begin{figure}[t]
\centering
\includegraphics[width=1.0\linewidth]{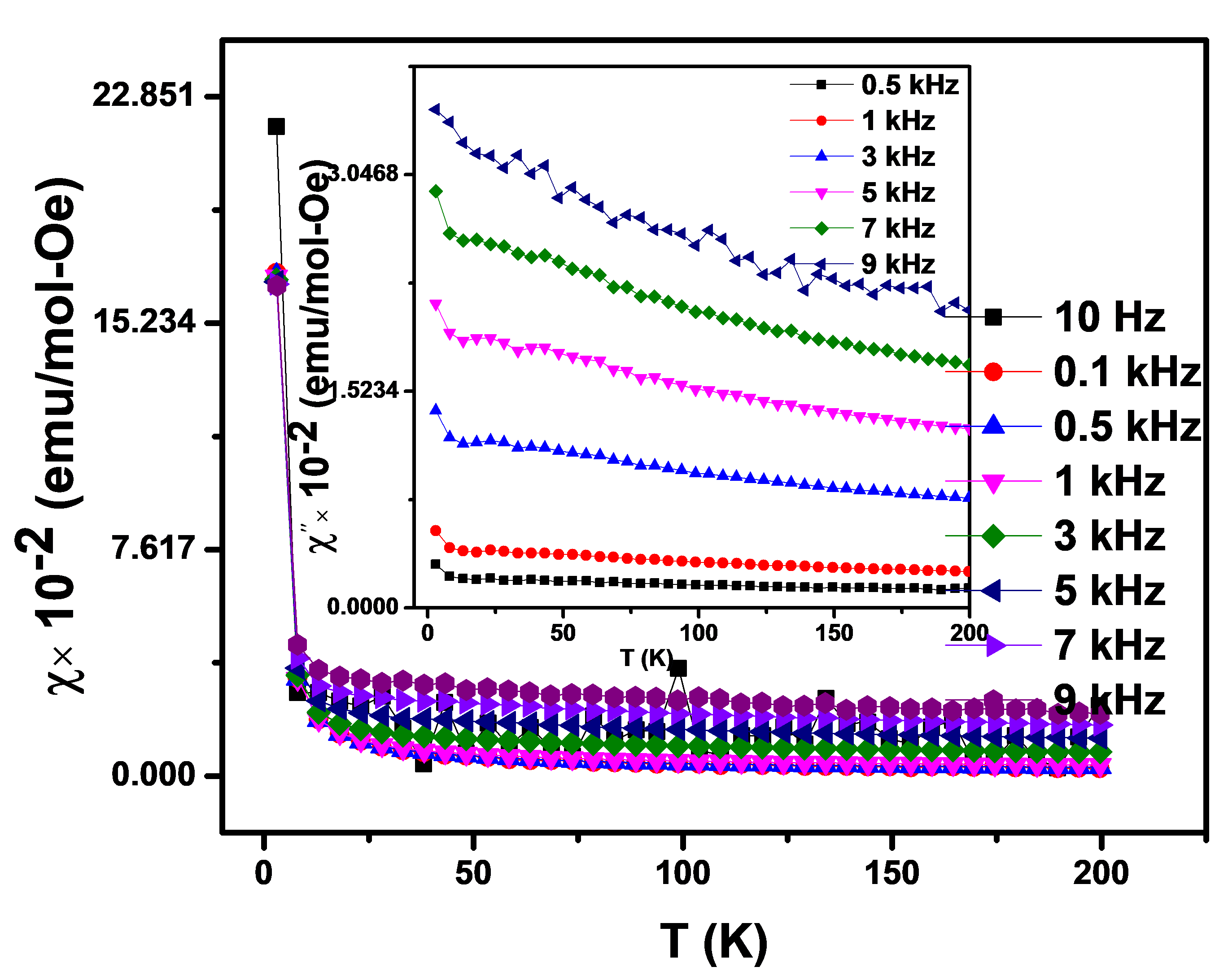}
\caption{(Color online) For CrFeVGa, T-variation of ac susceptibility for various frequencies driven at 5 Oe ac field. Inset shows T-variation of out of phase ACS ($\chi$$'$).}
\label{fig:ac}
\end{figure}

CFVG should possess 2.0 $\mu_B$/f.u. in the fully ordered state. But interestingly, the observed moment turns out to be negligibly small ($5\times 10^{-2}$ $\mu_B$/f.u.). The presence of B2 
disorder can be a plausible reason for the quenching of the moment. The susceptibility data (H=100 Oe) have been fitted (solid red line) above 145 K (C-W law deviates below 150 K) 
using the Curie-Weiss law ($\chi^{-1}=\frac{1}{\chi_0 + C/(T-\theta_P)}$). From the fitting, we obtained $\chi_0$=$3.74 \times 10^{-4}$ emu/mol-Oe  and Weiss temperature, $\theta_W$= $-$25 K. 
The latter indicates the presence of antiferromagnetic (AFM) interactions in the system. Non-saturating behavior (even up to 60 kOe field) of low-T M-H curves, along with negligible hysteresis 
and very low moment ($\sim 10^{-2}$ $\mu_B$/f.u.) indicate superparamagnetic (SPM)-like behavior, possibly attributable to the B2 disorder.

\begin{figure*}[t]
\centering
\includegraphics[width=1.0\linewidth]{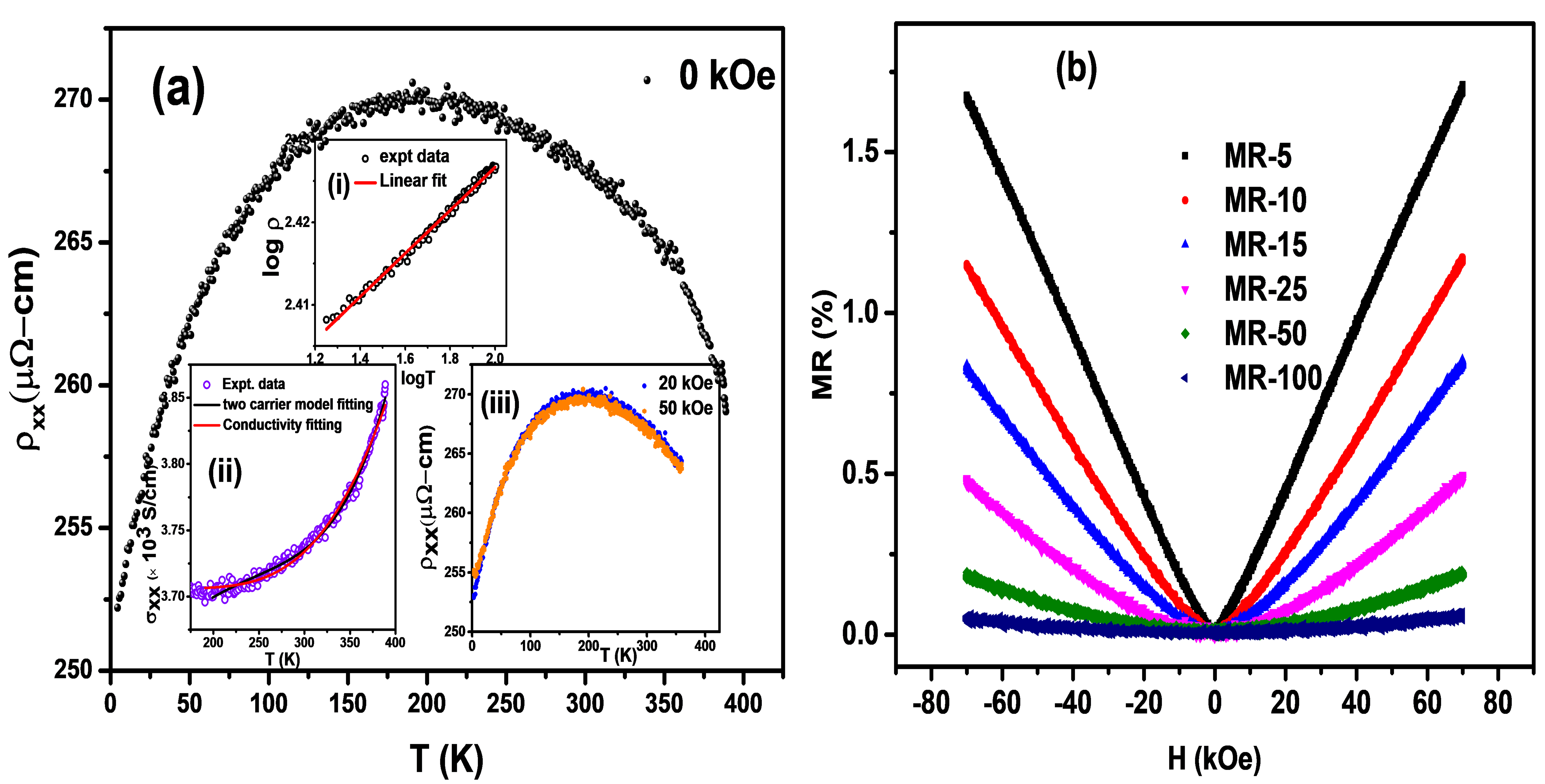}
\caption{ (Color online) For CrFeVGa, (a) Longitudinal resistivity ($\rho_{xx}$) vs. T in zero field. (Inset-i) $\log(\rho) $ vs. $\log(T)$ plot fitted in the T-range 18-100 K at 0 kOe (Inset-ii) longitudinal conductivity ($\sigma_{xx}$) vs. T in zero field along with a two-carrier model fit (Eq. \ref{eq:tbm-final}) and the conductivity fitting (Eq. \ref{eq:c}) in the semiconducting regime (200-390 K) (Inset-iii) $\rho_{xx}$ vs. T at two different fields. (b) $MR$ vs. $H$ at various temperatures in the range of 5-100K.}
\label{fig:RT-CFVG}
\end{figure*}

Furthermore, $1/\chi $ vs. T plot (see insets(i-ii) of Fig. \ref{fig:mt-CFVG}) depicts a deviation from the C-W behavior below 150 K for H=100 Oe and 100 K for H=50 kOe respectively, indicating 
the possibility of Griffith's phase (GP)-like behavior in this system.\cite{Karmakar_2013} We have defined the Griffith's temperature $T_G$ as the temperature at which $1/\chi$ deviates from the 
CW law. $T_G$ is the highest temperature at which there exists a short-range ferromagnetic (FM) ordering, while the system is completely paramagnetic above $T_G$. For H=100 Oe and H=50 kOe data, 
$T_G$ is 149 K and 100 K respectively. As the field is increased, $T_G$ decreases, which is a characteristic feature of GP\cite{Karmakar_2013}. This is because, at the lower fields, the moments 
of ferromagnetic (FM) clusters can easily prevail over the paramagnetic regime, while the same does not happen at higher fields.
Generally, GP is observed in frustrated magnetic systems,\cite{Karmakar_2013} and rarely seen in Heusler systems. The presence of anti-site disorder expedites the possibility of GP in this 
system, and in turn plays a significant role,\cite{PhysRevLett.88.197203,PhysRevLett.59.586} leading to interesting magnetic features. The competition between FM and AFM phases can also be a 
source of occurrence of GP. The existence of GP-like behavior in CFVG can be explained on the basis of quenched disorder and the coexistence of competing FM and 
AFM phases.\cite{PhysRevLett.59.586} Griffith's singularity\cite{griffiths1969nonanalytic} is represented by a power law of $1/\chi$ as,
 \begin{equation}
\chi^{-1} = {( T-T_{C}^{R})}^{1-\lambda}
\label{eq:Gp}
\end{equation}
here, $T_{C}^{R}$ is the random critical temperature and $\lambda$ is the susceptibility exponent (0$\leq$ $\lambda$ $\leq$1), which means the deviation from the C-W behavior. In the 
paramagnetic regime (above $T_G$), $\lambda$ should ideally be $0$.\cite{griffiths1969nonanalytic}

 Figure \ref{fig:GP}(i-ii) shows a fit to our $\chi^{-1}$ data (using Eq. \ref{eq:Gp}) to look into the presence of GP-like behavior. From the fitting, we obtained $\lambda$ $=0.88\pm0.05$, which 
is in good agreement with the expected range of 0-1 and comparable to that of the other reported systems from the Heusler family,\cite{dash2020structural,qian2021griffiths} indicating the 
presence of GP-like behavior. The calculated value of $T_{C}^{R}$=25 K (as $\theta_{CW}$ is negative in our case, and no magnetic ordering is found down to 2 K, we have approximated 
$T_{C}^{R}$=$| {\theta_{CW}}|$),\cite{Karmakar_2013} which is also above the ordering temperature. We found $\lambda$ $=0.88\pm0.05$ from the fitting of 100 Oe data, which is higher than that of 
50 kOe data ($\lambda$ $=0.79\pm0.02$ for 50 kOe). Thus $\lambda$ is higher in the lower field, again a characteristic feature of GP-like behavior.\cite{Karmakar_2013}
To further analyze this behavior in more detail, we have checked the Arrott plots along with a linear fit, as shown in Fig. \ref{fig:GP}(iii). The presence of non-linear behavior 
indicates the presence of magnetic inhomogeneities \cite{Karmakar_2013} and the absence of any spontaneous magnetization, which again confirms the presence of GP-like behavior.

To further investigate the magnetic properties of CFVG, we have performed frequency-dependent ACS measurements in the range of 3-200 K using ac frequencies of 10 Hz to 9 kHz. 
Figure \ref{fig:ac} shows the T variation of the real ($\chi$) and imaginary ($\chi$$'$) components of ACS, which agree with the DC magnetization data. The absence of a prominent peak confirms 
no magnetic ordering down to low T. The absence of any frequency dependence in the ACS data rules out spin-glass nature in this system. Thus, DC and AC magnetization data reveal the possibility 
of small magnetic clusters in the paramagnetic regime, formed by weakly interacting magnetic moments, where no spontaneous magnetization is observed (as revealed from the Arrott plot). 
This also confirms the absence of coherent long-range order in this system, because of the atomic disorder.

\subsection{Transport properties}
\subsubsection{Resistivity}
Fig. \ref{fig:RT-CFVG} (a) shows the T-dependence of longitudinal resistivity ($\rho_{xx}$) at zero field and at different fields (see inset (iii) of Fig. \ref{fig:RT-CFVG}(a)).
In the low-T region, it is a bad metal as $\rho_{xx}$ increases with T very slowly. This is also been verified by plotting $\log(\rho_{xx}) $ vs. $\log$(T) in the T-range $18-100$ K 
along with a perfect linear fitting (see inset (i) of Fig. \ref{fig:RT-CFVG}(a)). From the fitting, we obtained the value of exponent $\alpha$ to be 0.02, which is negligibly small, 
indicating the semimetallic nature of CFVG. On the other hand, the resistivity data shows a semiconducting-like behavior above a temperature $T_m$ ($\sim$200K) at which $\rho_{xx}$ reaches 
a maximum and this maximum could be due to the competition between positive and negative temperature coefficients of
resistivity and/or increase of phonon scattering with T, while the effective carrier density remains constant (this is also revealed by the electronic structure 
calculations shown later). This semimetallic-to-semiconductor-like behavior indicates a gapless/small-gap semiconductor/semimetal-like feature.\cite{chen2021large} 
From the variation of the resistivity data, it appears that in the high-T regime (200-390 K) semiconducting behavior dominates possibly due to the effect of thermally 
activated carriers.

\begin{figure}[t]
\centering
\includegraphics[width=1.05\linewidth]{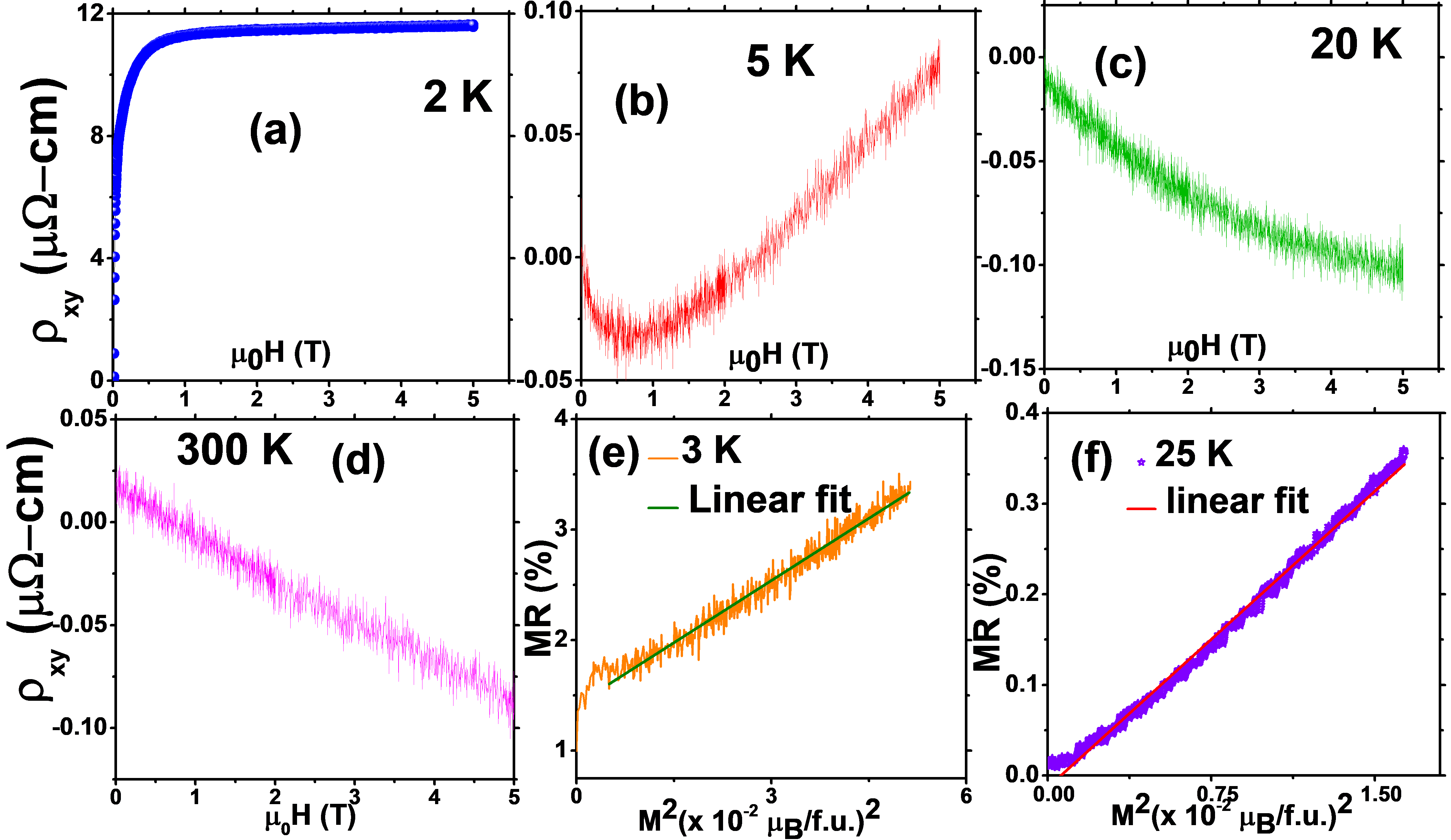}
\caption{(Color online) For CrFeVGa, (a-d) Hall resistivity ($\rho_{xy}$) vs. H at 2, 5, 20 and 300 K. (e-f) MR vs. M$^2$ with linear fit (solid line) at 3 and 25 K respectively.}
\label{fig:THall-CFVG}
\end{figure}

In order to further investigate the transport nature in the semiconducting regime, we have fitted the conductivity data with Eq. \ref{eq:c} (in the T-range $150-390$ K) and a modified 
two-carrier model, Eq. \ref{eq:tbm-final}\cite{kittel2007introduction,jamer2017compensated} (in the T-range $200-390$ K). This is shown in the inset (ii) of Fig. \ref{fig:RT-CFVG}(a)).
\begin{equation}
 \sigma (T)=\sigma_0 + \sigma_g e^{-E_{g}/k_{B}T}
 \label{eq:c}
\end{equation}
where, $\sigma_0$ and $\sigma_g$ are T-independent parameters, $E_{g}$ is the energy gap and $k_{B}$ is the Boltzmann's constant.
We have considered the two-carrier model, for which,
\begin{equation}
\sigma(T) = e (n_e \mu_e + n_h \mu_h)
\label{eq:tbm}
\end{equation}
where, $n_i=n_{i0}\ e^{-\Delta E_i/k_\mathrm{B}T}$($i=e,h$) are the electron/hole carrier concentrations. $\mu_i$ and $\Delta E_i$ are the mobilities and pseudogaps respectively.
Eq.(\ref{eq:tbm}) can be written as,
\begin{equation}
\sigma(T) = [A_e(T) \ e^{-\Delta E_e/k_\mathrm{B}T} + A_h(T) \ e^{-\Delta E_h/k_\mathrm{B}T}].
\label{eq:tbm-final}
\end{equation}

The fit with Eq. \ref{eq:c} yielded a value of $E_g$=0.18 eV, which corresponds to a narrow band gap semiconductor. From the two-carrier model fitting, the energy gaps ($\Delta E_i$) turn out to be 0.36 meV and 0.28 eV in the two channels, which is in line with the energy gap found from the fitting with Eq. \ref{eq:c}.  It appears that the atomic disorder has reduced the $E_{g}$ significantly. Thus, two-carrier model indicates the defect scattering dominated transport in CFVG. It should be noted that a semiconducting to semimetallic-like transition behavior has no magnetic origin, as there is no sign of magnetic transition at/around $200$ K in the magnetization data. As such, this may arise due to the change of gap near the Fermi level (E$_F$) (as one of the pseudo-gaps is very small) attributed to the disorder.

\subsubsection{Magnetoresistance} Another interesting feature of the present system is the observation of non-saturating, linear positive magnetoresistance in a wide T-range. 
Figure \ref{fig:RT-CFVG}(b) shows the MR vs. H at different T, where MR ratio is defined as MR(H)=$ \left[ \rho(H) - \rho(0)\right]/\rho(0)$ $\times 100\%$. MR increases with H in a 
perfectly linear fashion in the entire field range, and the low-temperature MR is still linear at 70 kOe. Such a LPMR is usually observed in gapped semiconductors. At 3 K, we have obtained $\sim$ 4$\%$ MR at 
H=70 kOe, which is remarkably high in a disordered system as compared to other reported systems. \cite{PhysRevB.69.245116,PhysRevB.103.104427} The slope of the linear region decreases 
with increasing T, and above 100 K, the magnitude of MR becomes very small ($\sim$ 0.01 $\%$ at 300 K). The occurrence of LPMR is indeed anomalous, as the MR(H) dependency is usually quadratic 
with H. The origin of such a feature is ambiguous, which may arise due to several reasons e.g. (i) disorder-mediated mobility fluctuations\cite{PhysRevLett.88.066602} (ii) quantum linear MR behavior in the zero/small-gap electronic band structure near E$_F$ \cite{PhysRevB.58.2788,xu1997large} etc. For CFVG, LPMR  possibly arise due to the second reason.\cite{PhysRevB.103.104427} \textcolor{black}{Such large nonsaturating LPMR can be quite promising for high-speed electronics and next-generation spintronic devices.} As the origin of LPMR depends majorly on the mobility and carrier concentration parameters, we have further investigated the Hall effect as described below.

\begin{figure}[t]
\centering
\includegraphics[width=9cm,height=7.5cm]{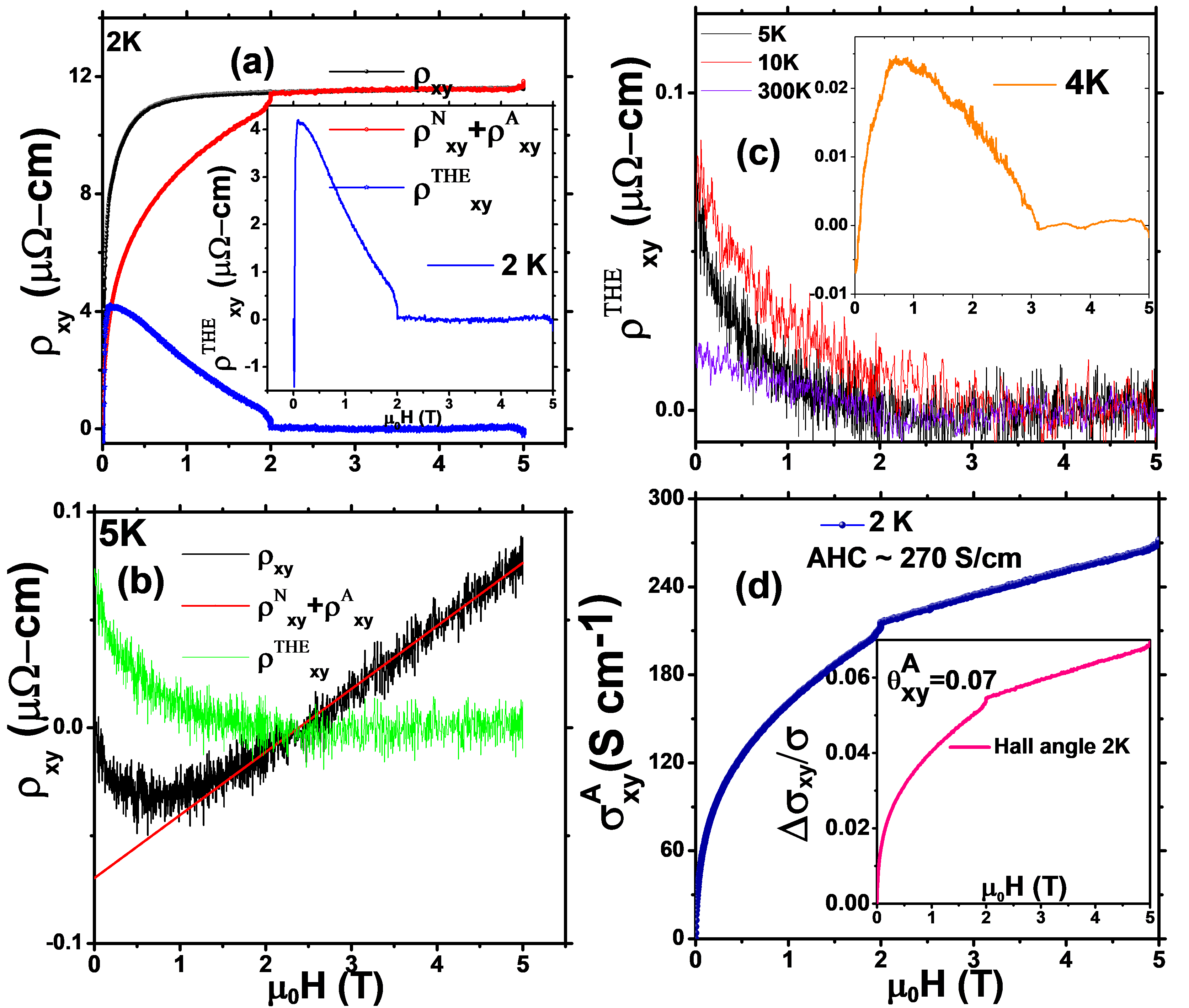}
\caption{(Color online) For CrFeVGa, (a) Total hall resistivity ($\rho_{xy}$) vs. H at 2 K (black curve). Red curve shows the sum of normal and anomalous Hall contribution, while the blue curve indicate topological Hall contribution (${\rho}^{THE}_{xy}$). Notice a change of sign in ${\rho}^{THE}_{xy}$ at $\sim$ 2T  (b) $\rho_{xy}$ vs. H at 5 K along with the normal, anomalous and topological Hall contributions. (c) ${\rho}^{THE}_{xy}$ vs. H at 5, 10, and 300K. Inset shows ${\rho}^{THE}_{xy}$ vs. H at 4K. (d) Anomalous Hall conductivity, AHC ($\sigma_{xy}^{A}$) vs. H at 2 K. Inset shows the anomalous Hall angle (AHA) vs. H.}
\label{fig:Hall-CFVG}
\end{figure}
\begin{figure}[t]
\centering
\includegraphics[width=1.0\linewidth]{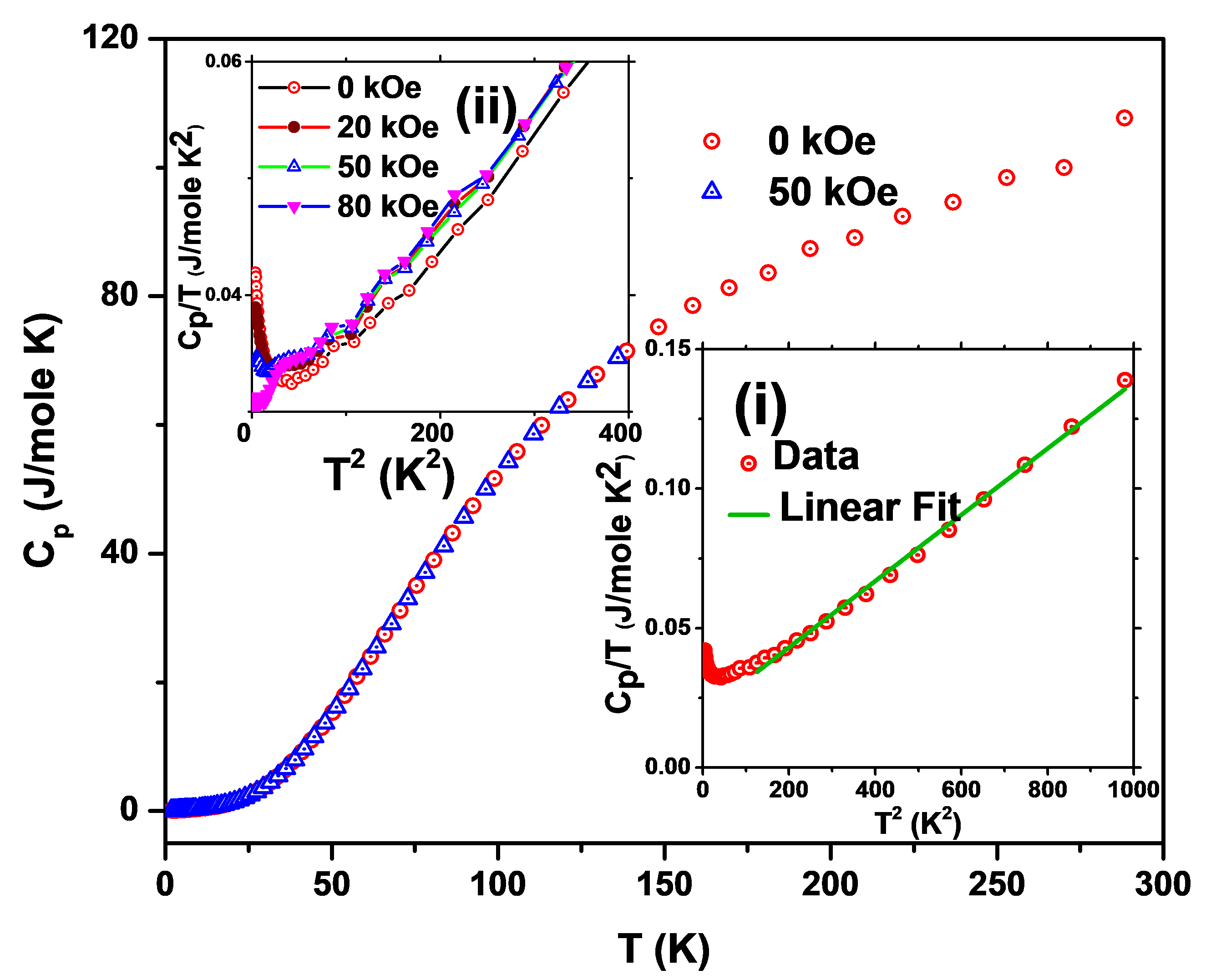}
\caption{(Color online) For CrFeVGa, specific heat (C$_p$) vs. T for 0, and 50 kOe fields. Insets (i) and (ii) show linear fit (solid green line) for 0 kOe data  and {C$_p$/T} vs. T$^2$ plot at various field values.}
\label{fig:hc}
\end{figure}

\subsubsection{Hall Measurements} 

\textcolor{black}{Interestingly, the Hall resistivities (MR contribution is removed from the Hall resistivity using the expression $\rho_{xy}$=$ \left[ \rho_{xy}(H) - \rho_{xy}(-H)\right]/2$) 
also show several anomalies (see Fig. \ref{fig:THall-CFVG} (a-d)) such as (1) hump-like maximum at 2 K, (2) non-linear behavior at low fields while linearity is observed at higher fields 
(3) switch of sign from negative to positive at low field and low-T and (4) a cross-over from p-type to n-type character. To better understand these anomalies, we have considered the empirical 
Hall resistivity expression for a magnetic material as,
\begin{equation}
\rho_{xy}=\rho_{xy}^{N} + \rho_{xy}^{A} =R_{0}H+R_{A}M,
\label{eq:Hall}
\end{equation}
where, $\rho_{xy}^{N}$, $\rho_{xy}^{A}$ are the normal and the anomalous Hall resistivity (AHE) contributions, $R_0$ and $R_A$ denote the ordinary and anomalous Hall coefficients respectively. 
AHE contribution can be scaled as $\rho_{xy}^{A}$=$S_{A}\rho_{xx}^{2}M$, $S_{A}$ is independent of field and $\rho_{xx}$ is the longitudinal resistivity. We experimentally observed a 
large anomalous Hall conductivity ($|\sigma_{xy}^{A}|$ = $\frac{\rho_{xy}}{\rho_{xy}^2+\rho_{xx}^2}$) of 270 S cm$^{-1}$ at 2 K, as shown in Fig. \ref{fig:Hall-CFVG} (d). 
We also estimated the carrier concentration $(n)$= 1.26$\times10^{19}$ cm$^{-3}$ at 2K, which is in the same order as that of semiconductor/semimetal carrier density, again indicating the 
semiconducting/semi-metallic nature of CFVG.\cite{chen2021large} To further quantify the AHE, we have used another scaling parameter, anomalous Hall angle (AHA), defined as, 
$\sigma_{xy}^{A}/\sigma_{xx}$$=(\sigma_{xy}-\sigma_{xy}^{N})/\sigma_{xx}$ and the maximum value of AHA reaches 0.07, as shown in the inset of Fig. \ref{fig:Hall-CFVG} (d). 
This AHA value is remarkably high among  bulk materials\cite{suzuki2016large}, as compared to other single crystal systems Mn$_3$Sn (AHA$\leq 0.02$) \cite{nakatsuji2015large} and in line 
with Mn$_3$Ge (AHA $\sim$ 0.05)\cite{nayak2016large}. However, $\rho_{xy}$ was found to deviate from the Hall resistivity fitting of Eq. \ref{eq:Hall}, as shown in Fig. \ref{fig:Hall-CFVG}(a-b). 
This clearly hints toward the existence of additional contribution in $\rho_{xy}$.\cite{PhysRevB.89.104408} The presence of hump-like maximum\cite{PhysRevLett.108.156601} and sign switching of the 
Hall resistivity\cite{doi:10.1063/5.0021722} are typical signatures of topological Hall effect (THE),\cite{PhysRevLett.110.117202} indicating the presence of THE in CFVG. 
Hence, in this case, total Hall resistivity can be expressed as
 \begin{equation}
\rho_{xy}(T)=\rho_{xy}^{N} + \rho_{xy}^{A}+  \rho_{xy}^{THE} =R_{0}H+S_{A}\rho_{xx}^{2}M+\rho_{xy}^{THE}
\label{eq:THE}
\end{equation}
where, $\rho_{xy}^{THE}$ is the topological Hall resistivity. To scale the AHE and THE contributions, we have performed a linear fit to $\rho_{xy}/H$ vs. $\rho_{xx}^{2}M/H$ curve in the 
high field regions, as THE vanishes at a high H value. From this fitting, R$_{0}$ and S$_{A}$ parameters are obtained, and THE is extracted by subtracting $\rho_{xy}^{N} + \rho_{xy}^{A}$ 
from Eq. \ref{eq:THE}\cite{PhysRevLett.110.117202}. This $\rho_{xy}^{THE}$ component is shown in Fig. \ref{fig:Hall-CFVG} (a-b). In addition, $\rho_{xy}^{THE}$ vs. H, at various T is 
shown in Fig. \ref{fig:Hall-CFVG} (c). From these fits, a large THE contribution ($\sim$ $4 \mu$ $\Omega$-cm) was observed at 2 K. The amplitude of THE decreases rapidly with T, but its 
contribution was found to exist over a wide T-range. The origin of this may be attributed to the existence of some spin texture\cite{PhysRevB.89.104408} and the non-trivial band topology near E$_F$, associated with 
the complex magnetic structure of this system mediated by atomic disorder. Furthermore, the presence of complex magnetic texture is also supported by the linear fit of MR vs. M$^{2}$\cite{doi:10.1063/5.0021722} (see Fig. \ref{fig:THall-CFVG}(e-f)). This concludes that CFVG exhibits an unconventional topological like 
Hall effect in addition to the conventional AHE.}


\textcolor{black}{
\subsubsection{Specific heat}
Figure \ref{fig:hc} shows the specific heat (C$_p$) vs. T for various applied fields. The low-T C$_p$ data has been fitted with the equation $C(T)=$ $\gamma$$T + $$\beta$T$^3$, where first 
and second term indicate electronic and low-T phonon contribution to C$_p$ respectively. Inset-(i) shows the linear fit of zero-field C$_p$ data while inset-(ii) shows {C$_p$/T} 
vs. T$^2$ plot. 
Sommerfeld coefficient, $\gamma$=0.0194 $J/mole-K^{2}$, is obtained from this fitting. We have then extracted the density of states (DoS) at Fermi level, n(E$_F$) $\sim$ 0.5 states/eV f.u., 
using the equation $n(E_F)$=$3\gamma/(\pi^{2}k_{B}^{2})$.\cite{venkateswara2019coexistence} This value matches fairly well with the simulated DoS value (described in next section). 
This is also in good agreement with small semimetallic DoS value near E$_F$ reported for other semimetals, and hence supports our transport and theoretical findings. From the fitting, the 
Debye temperature is found to be $\theta_D$= 254 K using $\beta$=1.188$\times$ $10^{-4}$ $J/mole-K^{4}$. Interestingly {C$_p$/T} vs. T$^2$ curve shows a shallow minima in low-T region, 
which vanishes with the field (see inset-(ii) Figure \ref{fig:hc}). The existence of a possible spin texture may be the reason for the anomalous low temperature 
behavior of C$_p$
}

\begin{figure}[t]
\centering
\includegraphics[width=1.0\linewidth]{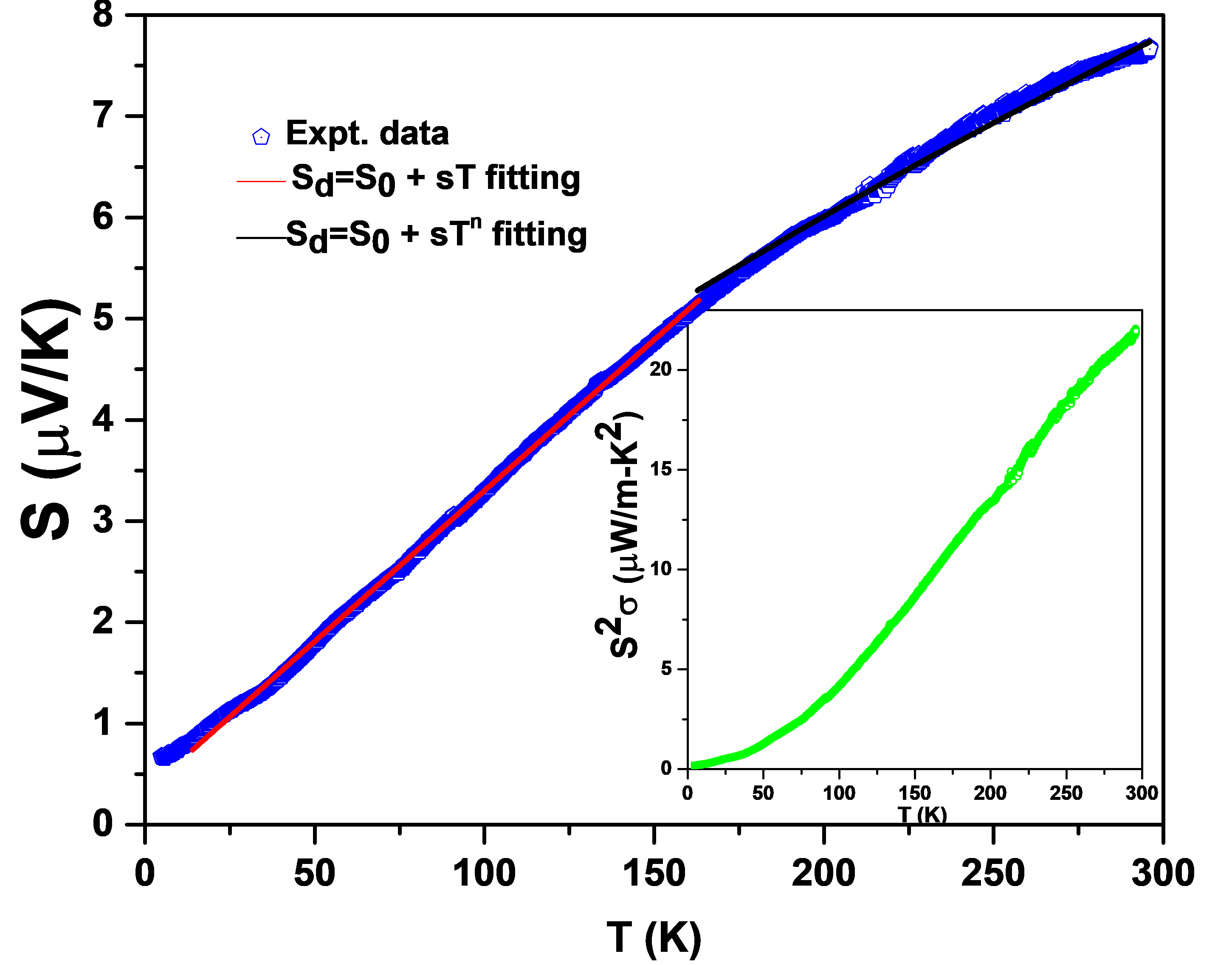}
\caption{(Color online) For CrFeVGa, thermoelectric power (S) vs. T along with a linear fit in the T-range of 20-150 K. Inset shows power factor (${S^2}\sigma$) vs. T.}
\label{fig:TEP}
\end{figure}

\subsubsection{Thermoelectric power (TEP)}
Figure \ref{fig:TEP} shows the T-dependence of the thermoelectric power (S) for CFVG. It shows a perfect linear dependence until 150 K, beyond which there is a slight change in the slope, 
which resembles the changeover in the resistivity data. The positive slope indicates a purely hole-driven thermoelectric power (TEP). The linear variation of S suggests a dominant 
contribution of diffusion thermopower, which is very similar to other reported narrow band gap semiconductors.\cite{PhysRevB.85.165149} The magnitude of S is in line with that of other 
reported semiconducting systems,\cite{PhysRevB.104.134406} but differs from that of the intrinsic semiconductors, where its magnitude is quite high. However, for  small/zero-gap semiconductors, 
E$_F$ can shift to the valence or conduction bands in one of the spin channels (which has a very small gap) due to excitation/impurity states. This leads to a low S value and a slow linear 
variation of S with T\cite{PhysRevB.104.134406}.

To find out the DoS and carrier density near E$_F$, TEP data have been fitted with the following equation in the two T-ranges (20-150 K and 150-300 K),

\begin{equation}
S_d=S_0+sT{^n}, 
\label{eq:S}
\end{equation}
where, $S_d$ is the diffusion thermopower, $S_0$ is a constant and slope $s$ = $\frac{\pi^2{k_B}^2}{3eE_{F}}$. From these fits, we obtained $E_F$ to be $0.8$ eV (20-150 K) and 0.2 eV (150-300 K) with $n=1.0$ and $n=0.7$ in the two T-ranges respectively. The inset of Fig. \ref{fig:TEP} shows the T-dependence of power factor ($S^2\sigma$).\cite{PhysRevB.103.085202}

\begin{figure}[t]
\centering
\includegraphics[width= 1.05\linewidth]{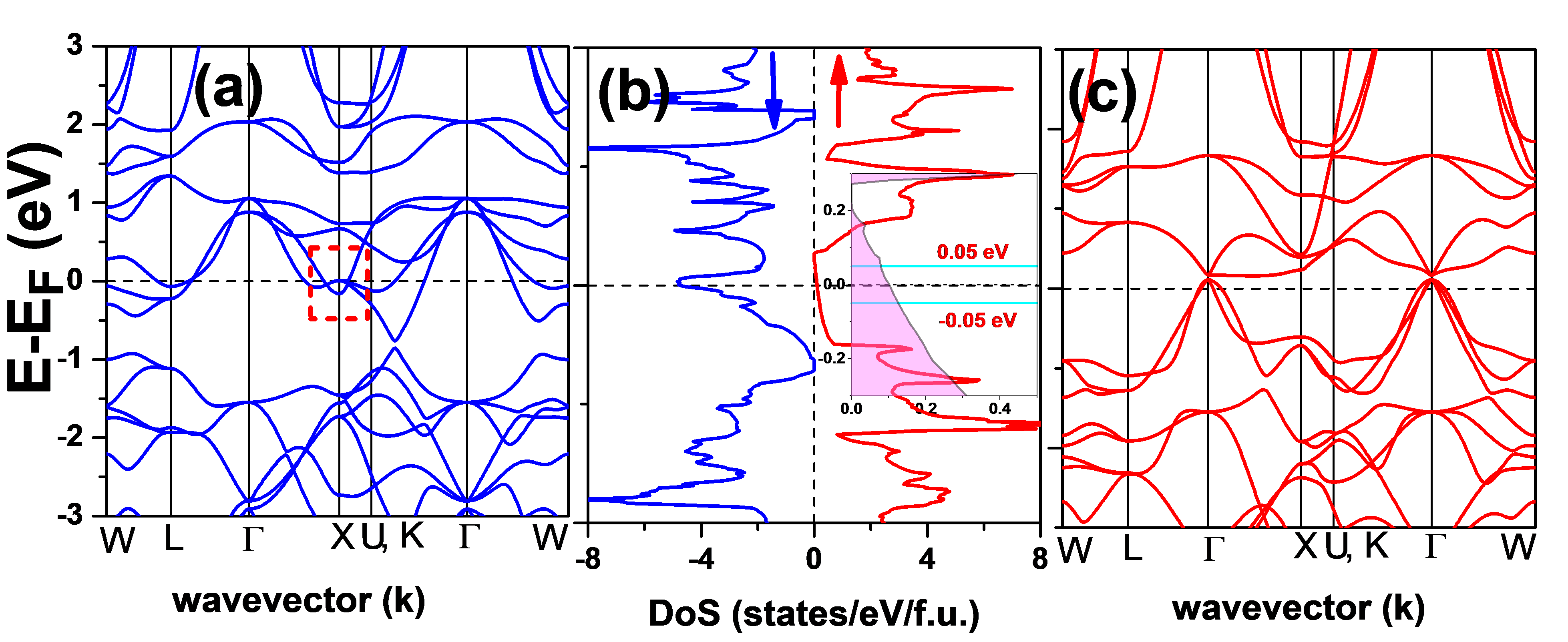}
\caption{(Color online) For CrFeVGa, spin polarized band structure and density of states at the relaxed lattice parameter($a_0$) for Type-I configuration with ferrimagnetic ordering. Red dashed rectangle in Fig. (a) highlights the topological non-trivial bands and Weyl node near the E$_F$ at/around X point. A zoomed in view of DoS near $E_F$ is highlighted in the inset of Fig. (b).}
\label{fig:CFVG-band}
\end{figure}

\section{Theoretical Results}
\begin{figure}[t]
\centering
\includegraphics[width=9cm,height=5cm]{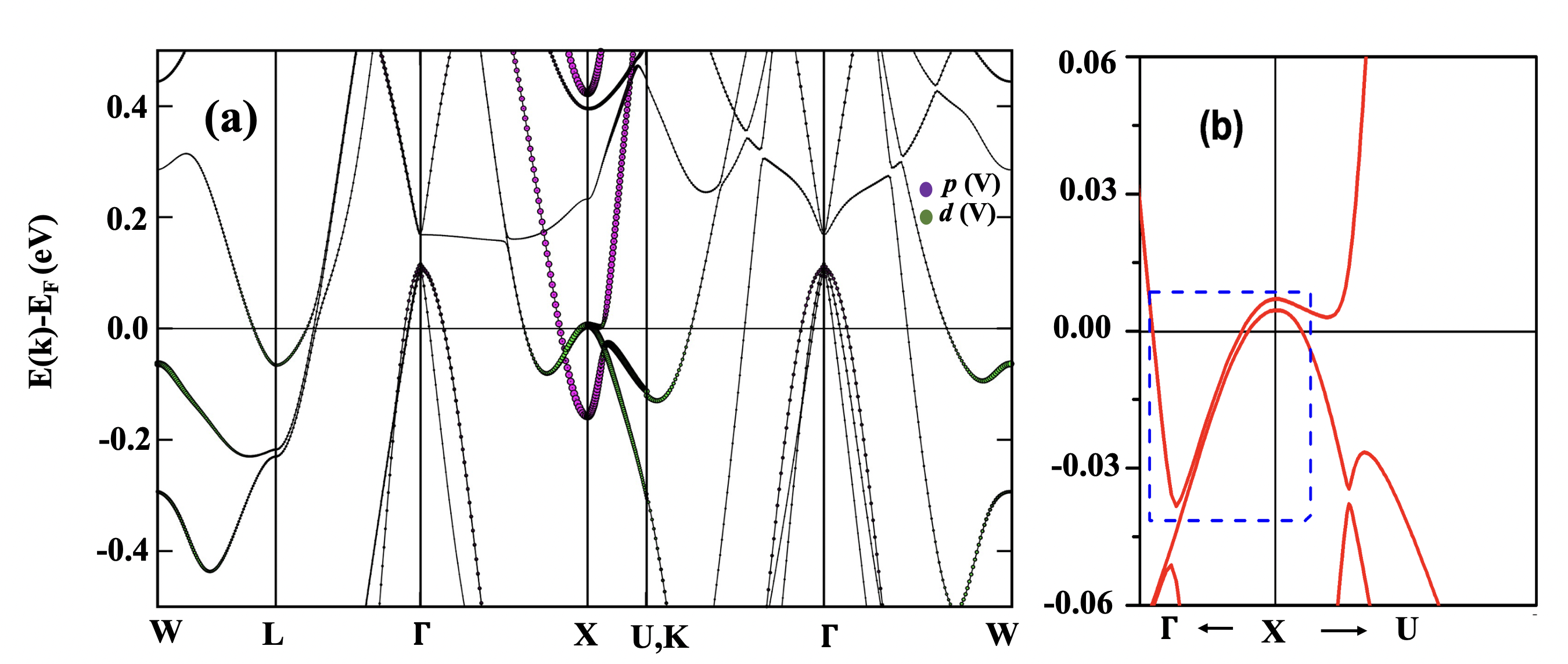}
\caption{(Color online) For CrFeVGa, (a) atom/orbital projected band structure including spin–orbit coupling(SOC) for Type-I configuration of CFVG. Vanadium p (d) states are shown by green(violet) color.(b) Zoomed-in view of the band structure around $X$-point highlighting the slight overlap of VB and CB near $E_F$, confirming the semimetallic nature of CFVG. }
\label{fig:soc}
\end{figure}

To study the electronic structure of CFVG, we have first simulated the energetics of various structural and magnetic states including ferro, antiferro-, and ferri-magnetic configurations. Of these, Type-I configuration with ferrimagnetic order turns out to be energetically the most stable one. Table \ref{tab:theory-CFVG} shows the simulated results for the theoretically optimized lattice parameters ($a_0$), atom-projected local moments and total energy for a few of the lowest energy configurations. Figure \ref{fig:CFVG-band} shows the spin-resolved band structure and density of states for Type-I configuration with ferrimagnetic disorder. A negligible overlap between the conduction and valence bands (CB and VB) and the presence of negligibly small DoS at $E_F$ indicate a weak semi-metallic behavior in this system. Interestingly, we observe the occurrence of topological non-trivial bands and Weyl points near E$_F$ around $X$ point, as highlighted by the red dashed rectangle in Fig. \ref{fig:CFVG-band}(a) (more details are given below). In the majority spin channel, small hole-pockets appear with a small band gap ($\sim$0.05 eV) very close to the E$_F$ at $\Gamma$ point.  Disorder can play a crucial role to push the E$_F$ towards VB/CB, which can impact the overall electronic structure of the material. It is worth mentioning that the DoS almost remains unchanged in the energy range of -50 meV to 50 meV, as shown in the inset of Fig. \ref{fig:CFVG-band}(b). This may arise due to the constant carrier density in the vicinity of E$_F$.\cite{chen2021large} Theoretically relaxed lattice parameter ($a_0=5.88$ \AA) matches fairly well with the experimental value ($a_0=5.87$ \AA). We have obtained a simulated net moment of $\sim$ $2\ \mu_B$/f.u., which does not follow the experimentally measured value. This difference can be attributed to the B2 disorder present in the system. To further investigate the semimetallic nature, we have performed band structure calculations considering the effect of spin-orbit coupling (SOC). Figure \ref{fig:soc}(a) shows the atom/orbital projected band structure of CFVG with SOC. There is a gap-opening and slight overlap between VB and CB near the high-symmetry point $X$ (see zoomed-in  Fig. \ref{fig:soc}(b)), which reconfirms the semimetallic nature of CFVG.

\begin{figure}[t]
\centering
\includegraphics[width= 1.0\linewidth]{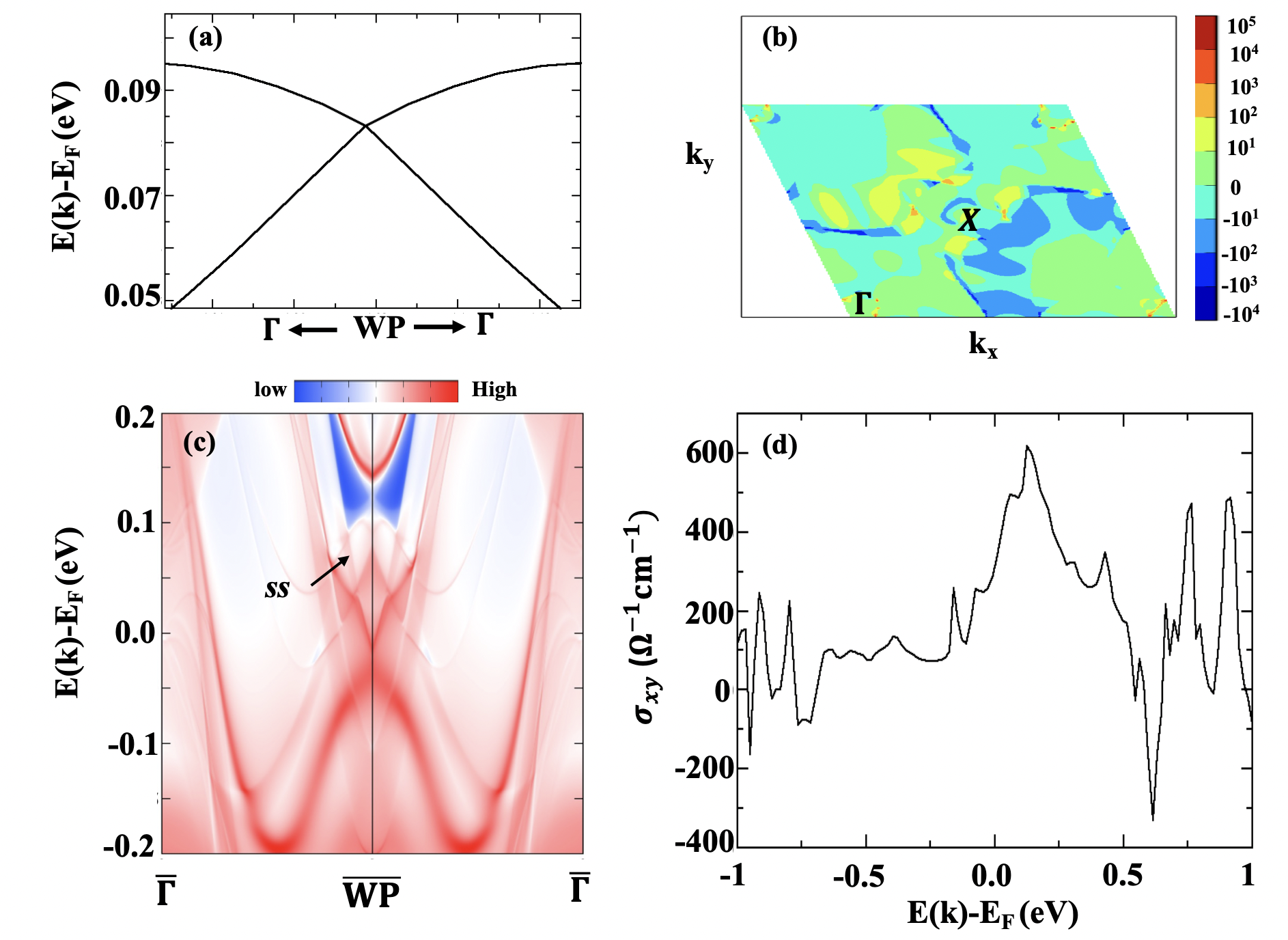}
\caption{(Color online) For CrFeVGa, (a) energy dispersion around a Weyl point (b) simulated Berry curvature in k$_x$-k$_y$ plane at the Fermi level (c) surface spectra around the Weyl point on (001) surface, $ss$ stands for surface state (d) energy-dependent anomalous Hall conductivity ($\sigma_{xy}$). }
\label{fig:tp2}
\end{figure}

\begin{table}[b]
\centering
\caption{For CrFeVGa, relaxed lattice parameter ($a_0$), atom-projected and total magnetic moments (in $\mu_B$), and relative energy ($\Delta E$) for type I, II and III configurations (with respect to type-I configuration) within the GGA approximation.}
\begin{tabular}{l c c c c c c}
\hline \hline
Type&  $a_0$ (\AA) $ \ $  &  $m^{\mathrm{Co}}$ & $\ $ $m^{\mathrm{Fe}}$ $\ $  &  $m^{\mathrm{V}}$  & $\ $ $m^{\mathrm{Total}}$ $ \ $ & $\Delta E$(eV/f.u.) \\ \hline
I   &  5.88   & 1.83 	& 	 0.88  	& 	 -0.75 	& 1.94	&  0   \\
II  &  5.85  & -0.96		& 1.29	& 	1.56 	&  1.91		& 0.54    \\
III    & 5.93  &  2.39  	&     1.98 	&	-0.5 	&  3.8		&  0.64   \\
\hline \hline
\end{tabular}
\label{tab:theory-CFVG}
\end{table}

The interesting band profile around the Fermi level inspired us to further inspect the topological properties of CrVFeGa. It is apparent that, combination of broken time-reversal symmetry and non-centrosymmetric nature of CFVG allow the possibility of occurrence of Weyl nodes in the bulk band structure. In order to examine the non-trivial band topology we have considered the `p' and `d' projected electronic band structure, as shown in Fig. \ref{fig:soc}.  In CFVG, the band inversion stems from the overlapping of Vanadium `p' and `d' states around X-point. The search of nodal points in the entire Brillouin zone revealed 24 pairs of Weyl points near the Fermi level with $\pm$1 chirality. The vanishing of net chirality value is in agreement with the Nielsen-Ninomiya theorem\cite{NIELSEN1981219}. Figure \ref{fig:tp2}(a) shows the band dispersion around one such Weyl point.  A highly linearized dispersive band nature is evident. \textcolor{black}{Such topologically non-trivial feature can be useful for several applications such as those in spin topological field effect transistors, broadband infrared photodetectors and topotronics.}
Further, we have examined the Berry curvature in k$_x$-k$_y$ plane, as shown in  Fig. \ref{fig:tp2}(b). Clearly, the magnitude of Berry flux is significantly high, which is possibly due to the existence of multi Weyl points near the Fermi level that acts as the source or sink of the Berry curvature. The surface states originating from the bulk nontrivial band crossing is one of the important parameters in topological materials, which are ideally protected against small external perturbations. Figure \ref{fig:tp2}(c) shows the surface state projected on the (001) surface originating from one of the Weyl points. \textcolor{black}{Such robust surface states can help to enhance the surface-related chemical processes of traditional catalysts.\cite{PhysRevLett.107.056804}}
In the vicinity of the Fermi level, one can observe the contributions of other surface states and arc, which arises from other Fermi pockets in the band structure of CFVG. Magnetic Weyl semimetals are well known for anomalous transport behaviour. In order to check that, we have calculated the intrinsic contribution of anomalous Hall conductivity originating from the topological Berry curvature.  Since the magnetization of the alloy is directed along the \emph{z-}direction, one need to analyze the transverse conductivity parameter ($\sigma_{xy}$). We have simulated the anomalous Hall conductivity using the following equation,

\begin{eqnarray*}
\sigma_{\alpha\beta} = -\frac{e^2}{\hbar}\int_{BZ}\frac{d^3k}{(2\pi)^3}\Omega_{\alpha\beta}(k)
\end{eqnarray*}

\begin{figure}[t]
\centering
\includegraphics[width= 8.5cm,height=6.0cm]{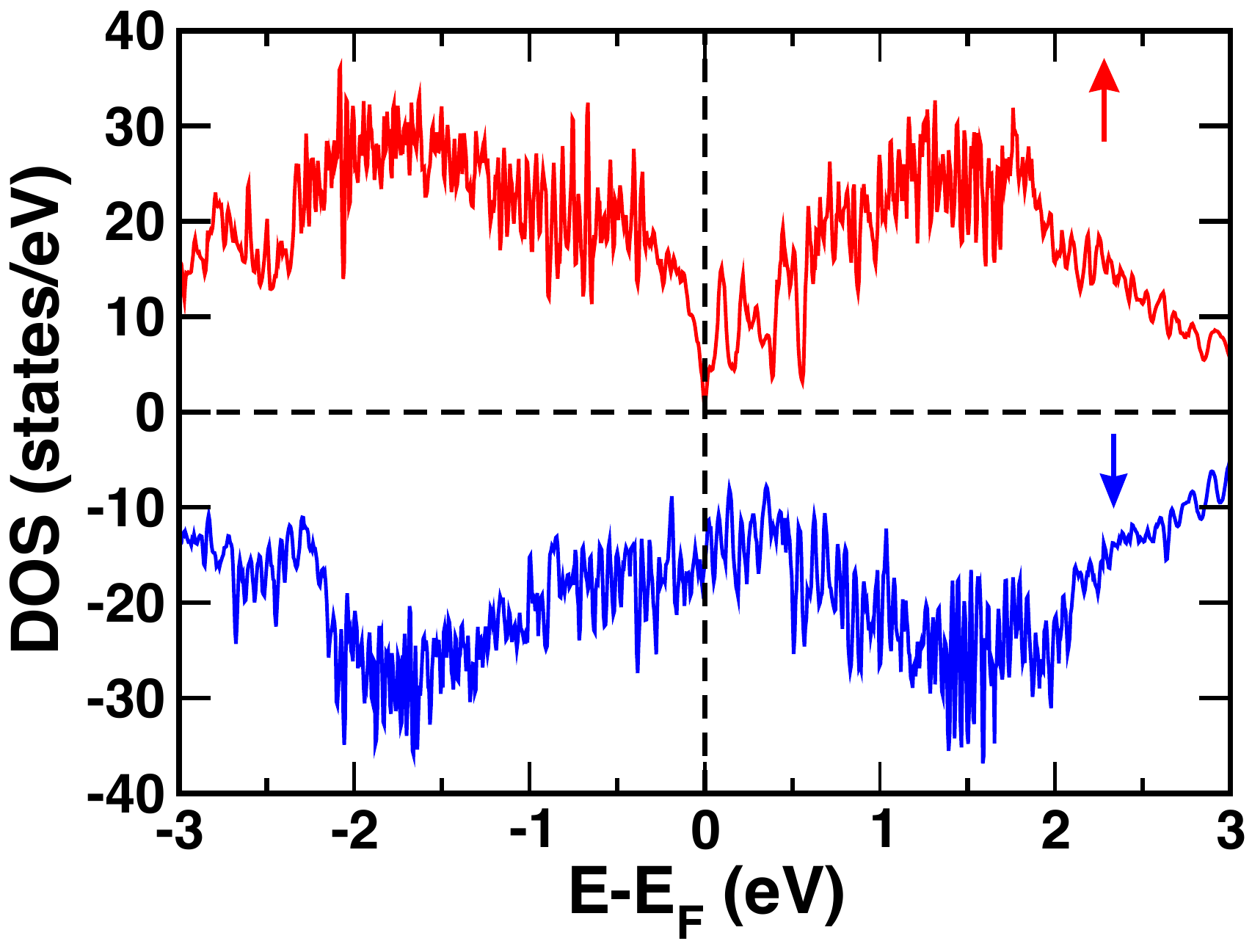}
\caption{(Color online) Spin polarized density of states for CrFeVGa with B2 disorder (i.e. 50\% disorder between Cr/Fe atoms and 50 \% disorder between V/Ga atoms). }
\label{fig:VNPicture2}
\end{figure}

Figure \ref{fig:tp2}(d) shows the calculated energy-dependent AHC for CFVG. As expected, the magnitude of AHC in the vicinity of the Fermi level is significantly high ($\sigma_{xy}(E_F)\sim$ 305 $\Omega$$^{-1}$  cm$^{-1}$), which is in good agreement with the experimentally obtained AHC value ($\sim$ 270 $\Omega$$^{-1}$  cm$^{-1}$). Further, along the positive side of energy, the AHC value is found to increase and attain a maximum value of 616 $\Omega$$^{-1}$  cm$^{-1}$ at 0.12 eV. Such a high AHC value for CFVG is comparable with other reported topological materials in the literature.\cite{guin20212d,kim2018jang} \textcolor{black}{and can be useful for different applications e.g. spin topological field effect transistors, topotronics etc.}

As mentioned earlier, magnetization measurements yield negligibly small moment for CFVG, which may arise due to the B2 disorder, as confirmed from our experimental XRD data.
In order to check it, we have simulated the effect of B2 disorder (fully homogeneous 50\% disorder between tetrahedral sites i.e. Cr/Fe atoms and octahedral sites i.e. V/Ga atoms) on the electronic/magnetic properties of CFVG. For this, we generated special quasi random structures(SQS)\cite{zunger1990special} corresponding to the energetically most favorable Type-I ordered configuration (see Table \ref{tab:theory-CFVG}).
Figure \ref{fig:VNPicture2} shows the spin polarized DoS for this SQS structure. Interestingly, the disordered phase also predict a semimetal behavior for CFVG, with finite DoS in one spin channel and almost zero gap in the other. Energetically, this phase differ only by a few meV/f.u. as compared to its corresponding ordered counterpart. The disordered phase however gives a much smaller net magnetization ($\sim$ 0.2 $\mu_B$), which goes in accordance with the experimental findings. Further, we have also simulated the electronic structure of B2-disordered phase using a 64-atom SQS unit cell. The DOS remains almost the same, but the net moment reduces to $\sim$ 0.11 $\mu_B$, in better agreement with the measured value.

\section{Summary and Conclusion}
In summary, we report a new topological semimetallic system CrFeVGa which belongs to quaternary Heusler alloy family. CrFeVGa crystallizes in a cubic structure with 50\% B2 disorder between 
tetrahedral sites i.e. Cr/Fe atoms and octahedral sites i.e. V/Ga atoms, as confirmed by the synchrotron XRD measurement. We use a combined theoretical and experimental study to investigate 
the effect of atomic disorder on the structural, magnetic, transport, electronic and topological properties of this alloy. B2 disorder is found to play a crucial role in the electronic/magnetic 
behavior of the system and possibly gives rise to quenching of moment ($\sim$ $5\times 10^{-2}$ $\mu_B$/f.u.) and other anomalies. AC and DC magnetization data reveal that the competition between 
antiferromagnetic phases with the small ferrimagnetic clusters help to mediate the Griffith's phase-like behavior, which eventually leads to anomalous magnetic transition. Resistivity data 
reflect disorder-mediated semiconducting to semimetallic transition in CrFeVGa. A non-saturating linear positive magnetoresistance (LPMR) is observed till 70 kOe field in a wide temperature 
range, which possibly originates from quantum linear MR feature in the zero/small-gap electronic band structure near E$_F$. Hall measurements support the transport data and display 
a few anomalies, which was further explained by the simulated band structure of the alloy. \textcolor{black}{Low-T anomaly in the specific heat data supports the possible existence of spin texture, as also observed in Hall measurements.} Ab-initio density functional calculations reveal the topological non-trivial features including band
inversion, Weyl points, and large Berry curvature for pristine CrFeVGa. The simulated value of intrinsic anomalous Hall conductivity of CrFeVGa matches fairly well with the experiment, 
and is found to originate mainly from the large Berry flux. Simulation of special quasi random structure (SQS) confirms the B2 disorder to be mainly responsible for the quenching of net 
magnetization.  The present study is crucial to get an insight into the effect of inhomogeneous phases on the critical behavior of weak magnetically ordered systems.The coexistence of so 
many emerging features such as \textcolor{black}{topological band structure, possible spin texture, large LPMR, high AHC value} in a single material is remarkable and it opens up new opportunities
for future topological/spintronics based research.


\section{Acknowledgment }
JN acknowledges the financial support provided by IIT Bombay. JN also thanks Dr. Velaga Srihari, ECXRD beamline, BL-11, Indus-2, RRCAT for carrying out anomalous XRD measurements. 
The authors thank Dr. Manoj Raama Varma, National Institute for Interdisciplinary Sciences and Technology (CSIR), Thiruvananthapuram, India for ACS measurements, and Dr. Durgesh Singh 
for setting up thermoelectric power measurements. JN and P.C.S thank IIT Bombay space time computing facility. KGS thanks DST-SERB (Grant No. CRG/2020/005589). AA acknowledges DST-SERB 
(Grant No. CRG/2019/002050) for funding to support this research.


\bibliographystyle{apsrev4-2}
\bibliography{references}

\end{document}